\documentclass[%
 aip,
 jcp,%
 amsmath,amssymb,
 reprint,%
]{revtex4-1}

\usepackage{graphicx}
\usepackage{dcolumn}
\usepackage{bm}
\usepackage{color} 
\usepackage{amsmath,amssymb}
\usepackage{epsfig}
\usepackage{bbold}
\usepackage{url}
\usepackage{lineno}
\usepackage[colorlinks=true,citecolor={blue},urlcolor={blue}, linkcolor=blue]{hyperref}%
\begin{document}
\title{Relativistic double-ionization equation-of-motion coupled-cluster method: Application to low-lying doubly ionized states}
\author{Himadri Pathak}%
\email{hmdrpthk@gmail.com}
\affiliation{Electronic Structure Theory Group, Physical Chemistry Division, CSIR-National Chemical Laboratory, Pune, 411\,008, India}%
\author{Sudip Sasmal}
\email{sudipsasmal.chem@gmail.com}
\affiliation{Electronic Structure Theory Group, Physical Chemistry Division, CSIR-National Chemical Laboratory, Pune, 411\,008, India}
\author{Kaushik Talukdar}
\email{talukdar.kaushik7970@gmail.com}
\affiliation{Department of Chemistry, Indian Institute of Technology Bombay, Powai, Mumbai 400\,076,  India}
\author{Malaya K. Nayak}
\affiliation{Theoretical Chemistry Section, Bhabha Atomic Research Centre, Trombay, Mumbai 400\,085, India}
\author{Nayana Vaval}
\affiliation{Electronic Structure Theory Group, Physical Chemistry Division, CSIR-National Chemical Laboratory, Pune, 411\,008, India}
\author{Sourav Pal}
\email{s.pal@iiserkol.ac.in}
\affiliation{Department of Chemistry, Indian Institute of Science Education and Research Kolkata, Mohanpur 741\,246, India}
\affiliation{Department of Chemistry, Indian Institute of Technology Bombay, Powai, Mumbai 400\,076, India}
\begin{abstract}
The article deals with the extension of the relativistic double-ionization equation-of-motion coupled-cluster (DI-EOMCC)
method [H. Pathak {\it et al.} Phys. Rev. A {\bf 90}, 010501(R) (2014)] for the molecular systems.
The Dirac-Coulomb (DC) Hamiltonian with four-component spinors is considered to take care of the relativistic effects.
The implemented method is employed to compute a few low-lying doubly ionized states of noble gas atoms
(Ar, Kr, Xe, and Rn) and Cl$_2$, Br$_2$, HBr, and HI. Additionally, we presented results with two intermediate schemes in the four-component relativistic DI-EOMCC framework
to understand the role of electron correlation. 
The computed double ionization spectra for the atomic systems are compared with the values from the 
non-relativistic DI-EOMCC method with spin-orbit coupling (SOC) [Z. Wang {\it et al.} 
J. Chem. Phys. {\bf 142}, 144109 (2015)] and
the values from the National Institute of Science and Technology (NIST) database.
Our atomic results are found to be in good agreement with the NIST values.
Further, the obtained results for the molecular systems agree well with the available experimental values.
\end{abstract}
\date{\today}
\maketitle
\section{Introduction}
There has been considerable progress in the last few decades in the experimental techniques for the study of di-cationic ions.
\cite{harris1992double, vekey1995multiply, lahmam1989electron, ackermann2007operation,sansone2006isolated, goulielmakis2008single}
The doubly charged ions are highly reactive, and they play an important role in ionized gases, excimer lasers, plasmas,
and interstellar clouds, which attracted much attention from both experimentalists and theoreticians. \cite{cox2003high, PhysRevA.51.R2668, prasad1975importance, 
rosner1998study}\par
The unambiguous theoretical prediction of the double-ionization spectra demands a reliable theory.
The correctness and authenticity of the calculated values depend on the quality of the wavefunction in describing the quantum systems.
Therefore, it is obligatory to have a highly correlated method that can provide a better description of the ionized states. Besides, electron correlation,
relativistic effects has a significant contribution to describe the simultaneous removal of two electrons.
Thus, concurrent treatment of both the effects of relativity and electron correlation
is mandatory due to their intertwined nature.\cite{grant2007relativistic, dyall2007introduction}\par
The self-consistent-field (SCF) solution considering Dirac-Coulomb Hamiltonian with four-component spinors provides
the best possible single determinantal description
of the effects of relativity in a quantum many-body calculations.
On the other hand, coupled-cluster theory\cite{bartlett2007coupled, crawford2000introduction, kummel2003biography} becomes
the most acceded many-body method while dealing with the electron correlation.
Therefore, the development of a relativistic coupled-cluster theory will be the means to engage with these problems.\par
Kaldor and coworkers extensively worked on the relativistic coupled-cluster theory.
\cite{ilyabaev1992relativistic, ilyabaev1992relativistic2, ilyabaev1993relativistic, visscher2001formulation, eliav1995transition, eliav1996element, PhysRevLett.118.023002}
They extended the Fock-space multireference coupled-cluster theory (FSMRCC)
\cite{lindgren1978coupled, haque1984application, stolarczyk1985coupled, pal1987multireference, lindgren1987connectivity, pal1988molecular, jeziorski1989valence, mukherjee1989use}
to the relativistic framework to calculate transition energies.\cite{ilyabaev1992relativistic2, eliav1995transition, eliav1996element, PhysRevLett.118.023002}
The effective Hamiltonian variant of the FSMRCC theory encounters convergence difficulty with an increase in the size of the model space.
The intermediate Hamiltonian Fock-space multireference
coupled-cluster (IHFSMRCC) theory\cite{malrieu1985intermediate, meissner1998fock, landau1999intermediate}
and the MRCC theory based on eigenvalue independent partitioning (EIP-MRCC)\cite{SINHA1989544,sceip} are
quite impressive in predicting spectroscopic properties in an intruder-free manner.\par
\begin{table*}[ht!]
\caption{ Basis information and correlation energies (in Hartree) from the MBPT(2) and CCSD method}
\begin{ruledtabular}
{%
\newcommand{\mc}[3]{\multicolumn{#1}{#2}{#3}}
\begin{center}
\begin{tabular}{lrrrrrr}
Atom/ & Basis & Virtual Cutoff & \mc{2}{c}{Spinor} & \mc{2}{c}{Correlation Energy}\\
\cline{4-5} \cline{6-7}
Molecule & × & (a.u.) & Occupied & Virtual & MBPT(2) & CCSD\\
\hline 
Ar     & dyall.cv3z\cite{dyall2016relativistic}& 500.0 & 18 & 146 & -0.59324132 & -0.60681938 \\
Kr     & dyall.acv3z\cite{dyall2006relativistic}& 500.0 & 36 & 230 & -1.46015195 & -1.39817620 \\
Xe     & dyall.acv3z\cite{dyall2006relativistic}& 500.0 & 54 & 242 & -1.62810633 & -1.53775344 \\
Rn     & dyall.acv4z\cite{dyall2006relativistic}& 200.0 & 86 & 364 & -2.38266005 & -2.16144919 \\
Cl$_2$ & dyall.av3z\cite{dyall2016relativistic}& 500.0 & 34 & 242 & -0.71772024 & -0.74447938\\
Br$_2$ & dyall.cv3z\cite{dyall2006relativistic}& 500.0 & 70 & 402 & -2.88914902 & -2.75863049 \\
HBr    & H: aug-cc-pVTZ\cite{dunning1989gaussian} &500.0 & 36 & 222 & -0.95277980 &-0.91704702 \\
       & Br: dyall.av3z\cite{dyall2006relativistic} &  &  &  &  & \\
HI     & H: aug-cc-pVTZ\cite{dunning1989gaussian} &500.0 & 54 & 308 & -2.27780739 &-2.16475927\\
       & I: dyall.ae3z\cite{dyall2006relativistic} &  &  &  &  &  \\
\end{tabular}
\end{center}
}%
\end{ruledtabular}
\label{basis}
\end{table*}
%

\begin{table*}[!t]
\caption{Comparison of DIP values (in eV) of Kr with various schemes in the EOMCCSD approximation}
\begin{ruledtabular}
\begin{center}
\begin{tabular}{llllll}
       & & &States&  &  \\
\hline       
Scheme &$^3P_2$& $^3P_1$ &$^3P_0$&$^1D_2$&$^1S_0$\\
\hline
basis: aug-cc-pVTZ\cite{wilson1999gaussian}                                      \\  
contracted basis, point nucleus, No of virtual spinor=82\\
\\
ECPDS10MDFSF\cite{peterson2003systematically}(spin free)&38.067($^3P$)&   &  &39.514($^1D$)&41.740($^1S$)\\ 
ECPDS10MDFSO\cite{peterson2003systematically}(spin orbit)&37.663&38.389&38.460&39.631&41.993\\
Dirac-Coulomb ($1s^22s^22p_{1/2}^22p_{3/2}^4$ core frozen)&38.203&38.670&38.775&40.005&42.316\\
Dirac-Coulomb                                             &38.202&38.669&38.774&40.004&42.315\\
\\
\\
basis: Dyall.av3z\cite{dyall2006relativistic} \\
finite nucleus, virtual cutoff =500 Hartree, No of virtual spinor=170\\
\\
Dirac-Coulomb&38.177&38.754&38.853&40.057&42.388 \\
Dirac-Coulomb-Gaunt&38.165&38.730&38.831&40.035&42.362\\
\\
basis: Dyall.acv3z\cite{dyall2006relativistic}                                \\
finite nucleus, virtual cutoff =500 Hartree, No of virtual spinor=230\\
\\
Dirac-Coulomb&38.341&38.930&39.028&40.218&42.566\\
Dirac-Coulomb-Gaunt &38.329&38.906&39.005&40.196&42.539\\
\\
basis: Dyall.aae3z\cite{dyall2006relativistic}                                 \\
finite nucleus, virtual cutoff =500 Hartree, No of virtual spinor=244\\
\\
Dirac-Coulomb&38.342&38.931&39.028&40.219&42.567\\
Dirac-Coulomb-Gaunt&38.330&38.907&39.006&40.196&42.540\\
\\
no virtual cutoff, No of virtual spinor=326 \\
\\
Dirac-Coulomb&38.344&38.933&39.031&40.224&42.569\\
Dirac-Coulomb-Gaunt&38.332&38.909&39.008&40.199&42.542\\ 
\\
NIST\cite{nist}    & 38.359&38.923&39.018&40.175&42.461
\end{tabular}
\end{center}
\end{ruledtabular}
\label{kr_example}
\end{table*}

As an alternative, the equation-of-motion coupled-cluster (EOMCC)
method\cite{eom1, eom2, eom3, eom4, eom5, eom6, eom7, eom8, eom9, eom10, eom11, eom12, eom13, eom14, eom15, eom16, eom17, eom18, eom19}
becomes popular due to
its simple CI-like eigenvalue structure,
hassle-free convergence, and the possibility of obtaining shake-up states those are important in explaining
photo-ionization spectra,\cite{ueda} and various radiation-less decay mechanism.\cite{nimrod}
Furthermore, the EOMCC method directly provides the eigenstates contrary to the propagator based approaches,\cite{linderberg2004propagators, cederbaum}
even though these methods share similar EOM structure.
%
We would like to emphasis that the EOMCC method works well at the noninteracting limit; however, it lacks the rigorous definition of the size-extensivity.
The values computed by the FSMRCC theory and the EOMCC method cease to agree with each other beyond the principal peaks of the one-valence sector.
Coupled-cluster linear response theory (CCLRT)\cite{monkhorst1977calculation, mukherjee1979response, koch1990coupled, koch1990coupled1} and symmetry adapted
cluster configuration interaction (SAC-CI) method\cite{nakatsuji1978cluster, nakatsuji1979cluster, nakatsuji1979cluster2}
are the alternative avenues for the calculation of direct energy differences closely related to the EOMCC method.
Chaudhuri {\it et al.} 
employed relativistic CCLRT within the four-component formalism
for the calculation of single ionization
potential of closed-shell atomic systems.\cite{chaudhuri1999relativistic,chaudhuri2000ionization}\par
It is desirable to introduce the effects of relativity in the electronic structure calculations by choosing an appropriate
relativistic Hamiltonian.
The use of relativistic effective-core-potential (RECP) with spin-orbit coupling (SOC) is the most common in molecular relativistic calculations.
A wide range of RECPs are available,
depending on how the RECPs are optimized. \cite{dolg2000effective, nicklass1995ab, dolg2012relativistic, schwerdtfeger2011pseudopotential}
It helps to exclude a large number of chemically inert electrons from the SCF calculations to reduce
the computational costs for the correlation calculation as compared to the fully relativistic counterparts.
%
%
The effects of SOC has been introduced
in different variants of the EOMCC method.
\cite{hirata2007high, epifanovsky2015spin, asthana2019exact, cao2017coupled, cheng2018perturbative, cao2016spin, wang2015equation, wang2014equation, tu2012equation}
However, the inclusion of the SOC with RECP does not address the intricate coupling between the relativistic and correlation effects.
\cite{dyall2007introduction, jensen2017introduction}
%
Therefore, it is essential to have a more robust theory considering relativistic Hamiltonian
with four-component wavefunction and a highly correlated method for the treatment of the electron correlation.
Further, we would like to categorically point out that the SOC effect is naturally taken care of by the Dirac-Hamiltonian.\cite{sakurai1967advanced}
The use of Dirac-Coulomb Hamiltonian is most common in relativistic electronic structure calculation
where the two-body Coulomb operator is added with the one-body Dirac-Hamiltonian.
The mathematical form of the Coulomb operator is the same as in the non-relativistic theory; however,
the physical meaning is different as it takes care of the spin-same orbit
interaction. The relativistic Hamiltonian containing up to Coulomb term is sufficient
for almost all chemical purposes.\cite{visser1992relativistic, saue2011relativistic} 
However, if unprecedented accurate results are sought especially for the fine-structure splitting from the deep core orbitals,
in such a case consideration of the spin-other-orbit interaction and spin-spin interaction become
relevant which require full inclusion of the Breit part of the two-body interaction.\par
The four-component relativistic EOMCC method has been implemented for the computation of 
ionization potential,\cite{pathak2014relativistic} electron affinity,\cite{blundell2014calculation}
and excitation energies,\cite{sahoo} of closed-shell heavy atomic systems, as well as for highly charged ions.\cite{pathak2015relativistic}
The atomic relativistic calculations enjoy the exploitation of the spherical symmetry,
which permits separate computation of the radial and angular part to work with the numerically evaluated reduced matrix elements.
On the other hand, in the spherical atomic implementation, the use of the antisymmetrized two-body matrix elements is not feasible due to the appearance of
the different angular factors for the direct and exchange part of the two-body matrix elements.
Thus, this non-separability of radial and angular part in the non-spherical case
makes molecular relativistic calculations onerous.
The EOMCC methods for the calculations of ionization potentials,\cite{pathak2014relativisticmol} and electron affinities\cite{pathak2016relativisticea}
of molecular systems considering both
four-component, as well as exact two-component (X2C) formalism\cite{liu2010ideas, saue2011relativistic} have been implemented and shed light
on the non-additivity of the relativistic effects and electron correlation effects through calculations.\cite{pathak2014relativisticmol}
Further, we have implemented the open-shell reference four-component EOMCC method and applied to calculate ionization potential of super-heavy
atomic and molecular systems using DC Hamiltonian.\cite{pathak2016relativisticopen}
Recently, Shee {\it et al.}\cite{shee2018equation} used Dirac-Coulomb-Gaunt Hamiltonian in their implementation of the EOMCC method.\par
The simultaneous removal of two electrons is a serious multireference problem. The
EOMCC method for double ionization potentials (DIPs),\cite{nooijen1997similarity, nooijen2002state, sattelmeyer2003use, demel2008application, musial2011multireference,
shen2013doubly} and double electron affinities (DEAs)\cite{shen2013doubly, musial2011multireference1, musial2014equation}
has been developed to deal with the complex multireference problem
within a single-reference description. However, those works are in the non-relativistic framework.
Further, we have implemented four-component relativistic DI-EOMCC method for closed-shell atomic systems and employed to calculate valence 
DIP values of alkaline earth metal atoms where valence electrons are well separated from the other inner electrons.\cite{pathak2014relativisticdip}\par
In this work, we extend the four-component DI-EOMCC method using DC Hamiltonian\cite{pathak2014relativisticdip} based on
antisymmetrized actual two-body matrix elements applicable to both atomic and 
molecular systems starting from their closed-shell configuration. We have employed to calculate a
few low-lying doubly-ionized states of noble gas atoms (Ar, Kr, Xe, Rn) and molecular systems (Cl$_2$, Br$_2$, HBr, and HI).

%
The ground state reference wavefunction is defined at the coupled-cluster single- and double- excitation level (CCSD) and the EOM matrix constructed 
in the 2$h$ and 3$h$-1$p$ space. 
Further, two intermediate schemes have been designed to analyse the roles of correlation contributions; 
one uses the ground state description of the second-order many-body perturbation theory [MBPT(2)], and in the later
the EOM matrix is constructed only in the 2$h$ space.\par
The outline of this paper is as follows. A brief description of 
the DI-EOMCC method is presented 
in Sec. \ref{sec2}. Sec. \ref{sec3} and Sec. \ref{sec4} are allocated for details of the computational parameters
and about the discussion of the obtained results in our calculations, respectively.
Finally, we convey our concluding thoughts in Sec. \ref{sec5}.
Atomic units are consistently used unless otherwise stated.

\section{Method} \label{sec2}
The wavefunction in the EOMCC method is defined as $Re^T|\Phi_0\rangle$,
where $e^T|\Phi_0\rangle$ is the coupled-cluster ground state wavefunction and $|\Phi_0\rangle$  
is the restricted closed-shell reference determinant.
$T$ is the usual cluster operator and the $R$ is a linear-operator.
The $R$ operator acts upon the coupled-cluster ground state wavefunction and generates the excited state configurations.
The second-quantization form of the cluster-operator $T$ and the EOM operator $R$ is as follows, 
\begin{equation}
\begin{split}
T=& T_{1}+T_{2}+\dots\\
             =&  \sum\limits_{{i,a}} t_{i}^{a}a_{a}^{\dag}a_{i} +
     \sum\limits_{\stackrel{a<b}{i<j}}t_{ij}^{ab}a_{a}^{\dag}a_{b}^{\dag}a_{j}a_{i}+\dots .
\end{split}
\label{tamp}
\end{equation}
\begin{equation}
\begin{split}
R=& R_{2}+R_{3}+\dots\\
             =&  \sum_{i<j} r_{ij} a_i a_j +  \sum\limits_{\stackrel{a}{i < j<k}} r_{ijk}^{a} a^{\dag}_a a_k a_j a_i+\dots ,
\end{split}
\end{equation}

$i,j, k \dots$($a, b, c, \dots$) are hole (particle) indices
that are occupied and unoccupied in the reference determinant, respectively. 
We have restricted our DI-EOMCC implementation up to $T=T_1+T_2$ and $R=R_2+R_3$ excitation level. 
The first step for the EOMCC calculation is the solution of the cluster amplitudes in Eq. \ref{tamp}. These amplitudes are obtained by iterative solution
of the following non-linear simultaneous equations, 
%
\begin{equation}
{\langle \Phi_{i}^{a}|e^{-T} He^{T}|\Phi_{0}\rangle=0,\,\,\,\,\, \langle \Phi_{ij}^{ab}|e^{-T} He^{T}|\Phi_{0}\rangle=0 }.
\end{equation}
where $|\Phi_{i}^{a}\rangle$ and $|\Phi_{ij}^{ab}\rangle$ are the single and double excited configurations with respect 
to the reference determinant. The ground-state energy obtained by solving equation for the energy,
\begin{equation}
E=\langle \Phi_{0}|e^{-T} He^{T}|\Phi_{0}\rangle , 
\end{equation}
Here, $H$ is the Dirac-Coulomb Hamiltonian and is of 
the follwing form,
\begin{eqnarray}
{H_{DC}} &=&\sum_{A}\sum_{i} [c (\vec {\alpha}\cdot \vec {p})_i + (\beta -{\mathbb{1}_4}) m_0c^{2} + V_{iA}] \nonumber\\
       &+& \sum_{i>j} \frac{1}{r_{ij}} {\mathbb{1}_4},
\end{eqnarray}
$\alpha$ and $\beta$ are the usual Dirac matrices. $V_{iA}$ is the potential energy operator
for the $i^{th}$ electron in the field of nucleus $A$.
$m_0c^2$ is the rest mass energy of the free electron, where $c$ stands for the speed of light.
The energy as well as the $R$ operator is determined by solving the following equation:
\begin{eqnarray}
\bar H R |\Phi_{0}\rangle=E R |\Phi_{0}\rangle
\end{eqnarray}
where $\bar H=e^{-T}He^{T}$ and $E$ is the energy of the doubly ionized state. 
\begin{table*}[!ht]
\caption{ Experimental and theoretical DIP values (in eV) in EOMCC approximation}
\begin{ruledtabular}
\begin{center}
\begin{tabular}{llrrrrr}
System &\, State &\,Ref. \cite{wang2015equation}&\, MBPT(2) &\, 2h &\,CCSD &\, NIST \cite{nist}\\
\hline
Ar&\,$^{3}P_2$&\,     &\,43.481&\,48.955&\,43.448&\,43.389\\
  &\,$^{3}P_1$&\,     &\,43.629&\,49.129&\,43.596&\,43.527\\
  &\,$^{3}P_0$&\,     &\,43.690&\,49.191&\,43.657&\,43.584\\
  &\,$^{1}D_2$&\,     &\,45.307&\,50.710&\,45.241&\, 45.126\\
  &\,$^{1}S_0$&\,     &\,47.793&\,51.771&\,47.694&\, 47.514\\
\hline
   \\
Kr & $^{3}P_2$ & 38.657 &38.558 & 42.880& 38.341 & 38.359\\
   & $^{3}P_1$ & 39.211 &39.149& 43.572& 38.930& 38.923\\
   & $^{3}P_0$ & 39.308 &39.248& 43.552& 39.028& 39.018\\
   & $^{1}D_2$ & 40.466 &40.462& 44.835& 40.218& 40.175\\
   & $^{1}S_0$ & 42.791 &42.811& 46.127& 42.566& 42.461\\
\hline   
   \\
Xe & $^{3}P_2$ & 33.406      & 33.460&36.742& 33.016& 33.105\\
   & $^{3}P_1$ & 34.597      & 34.725&38.202& 34.268& 34.319\\
   & $^{3}P_0$ & 34.413      &  34.514&37.586& 34.065& 34.113\\
   & $^{1}D_2$ & 35.516      &35.698&39.187& 35.222&35.225\\
   & $^{1}S_0$ & 37.897      &38.125&41.017& 37.659&37.581\\
\hline   
   \\
Rn & $^{3}P_2$ &  29.972     &30.339 & 33.070 & 29.837& 32.149(1.9)\\
   & $^{3}P_1$ &  33.814     &34.434 & 37.790 & 33.887 & \\
   & $^{3}P_0$ &  31.329     & 31.713 & 34.077 & 31.215 & \\
   & $^{1}D_2$ &  34.573     &35.203 & 38.578& 34.641 & \\
   & $^{1}S_0$ &  39.276     &40.146 & 43.773& 39.557 & \\
\end{tabular}
\end{center}
\end{ruledtabular}
\label{atomic}
\end{table*}

The above equation is projected onto the set of excited determinants (($|\Phi_{ij}\rangle$) and ($|\Phi_{ijk}^{a}\rangle$)
to obtain the following equations,
\begin{equation}
{\langle \Phi_{ji}|[\bar{H}, R_\nu]|\Phi_{0}\rangle=\Delta E_\nu r_{ji}},
\label{r1}
\end{equation}
\begin{equation}
{\langle \Phi_{kji}^{a}|[\bar {H},R_\nu]|\Phi_{0} \rangle =\Delta E_\nu r_{kji}^{a}},
\label{r2}
\end{equation}
Where $\Delta E_\nu$ is the amount of energy required to expel two electrons simultaneously from any given reference configuration. 
The commutative property of the $T$ and $R$ is assumed in deriving the above  equations.
The algebraic expression of the left hand sides of Eq. \ref{r1} and Eq. \ref{r2} are as follows,
\begin{widetext}
\begin{eqnarray}
\Delta E_\nu r_{ji}=&-&\hat P(ij)\sum_k \bar f_j^kr_{ki}
+0.5\sum_{l,k} \bar V_{ji}^{lk}r_{lk}+\sum_{k,a}\bar f_{a}^k r_{kji}^{a} 
-0.5\hat P(ij)\sum_{l,k,a} \bar V_{aj}^{lk}r_{lki}^{a}
\,\,\,\,\,\,\,\,\,\,\,\, \forall \,\,j<i
\end{eqnarray}
\begin{eqnarray}
\Delta E_\nu r^{a}_{kji}&=&
-\hat P(i|jk)\sum_l \bar V_{kj}^{al}r_{li}
-\hat P(ij|k)\sum_l \bar f_{k}^{l}r_{lji}^{a}
+\sum_{b} \bar f_{b}^{a}r_{kji}^{b} 
+ 0.5\hat P(i|jk)\sum_{l,m}\bar V_{kj}^{ml}r_{mli}^{a} 
-\hat P(ij|k)\sum_{l,b}\bar V_{kb}^{al}r_{lji}^b \nonumber \\
&+& 0.5\hat P(i|jk)\sum_{l,m,b}t_{kj}^{ab}\bar V_{ib}^{lm}r_{lm}
-0.5 \hat P(i|jk)\sum_{l,m,b,c}t_{kj}^{ac} V_{bc}^{lm}r_{lmi}^{b} \,\,\, \forall\,\,\, (a,\,\,k<j<i)
\end{eqnarray}
\end{widetext}
Here $\bar f $, $\bar V$ and $t_{kj}^{ac}V_{bc}^{lm}$  stand for one-body, two-body and three-body intermediate matrix elements
constructed by contracting appropriate one-body and two-body Hamiltonian matrix elements
and the converged amplitudes from the coupled-cluster ground-state calculation
as described in the Ref. \cite{shavitt2009many} $\hat P(i\dots|j\dots)$ stands for the cyclic permutation operator. 
The above equations can be expressed in the matrix form as $\bar H R= R\Delta E_\nu$.
The size of the $\bar H$ matrix is large enough ($nh^2+nh^3np, nh^2+nh^3np$, $nh$ and $np$
stand for the number of holes and particles, respectively)
to follow a full diagonalization algorithm in a reasonable basis.
Therefore, we have used Davidson diagonalization algorithm \cite{davidson197514} for the diagonalization of the non-Hermitian matrix.
The DI-EOMCC method is prone to slow convergence. 
Therefore, to obtain a smooth and faster convergence, we have used eigenvectors obtained from the full diagonalization of 2$h$
block as an initial guess for the iterative procedure.
The intermediate scheme MBPT(2) approximates CCSD ground state wavefunction at the second-order 
many-body perturbation theory level and for the $2h$ scheme, the dimension of the EOM matrix is restricted to ($nh^2, nh^2$), by ignoring
contribution from the $3h-1p$ block. 

\section{computational considerations}\label{sec3}
The SCF solution using DC Hamiltonian and the required one-body and two-body matrix elements for the correlation calculations
are obtained from the DIRAC14 program package.\cite{sauedirac}
Consideration of finite size nuclear model is most suited for the relativistic electronic structure calculations.
Therefore, the Gaussian charge density distribution nuclear model is taken into account to mimic the effects of the finite size nucleus.
All the parameters for this model are taken as default.\cite{visscher1997dirac}
The contribution from the high-lying orbitals in a correlation calculation is inconsequential due to their large energy values. 
Therefore, we have restricted the number of virtual orbitals on the basis of energy criteria. The orbitals above a threshold value
are discarded from the correlation calculations. 
%
The details of the basis set, threshold energy cutoff for the virtual orbitals including the number of occupied and virtual spinors are reported in Tab. \ref{basis}.
The two-body matrix elements below $10^{-12}$ are neglected in all our calculations due to their negligible
contribution in the correlation calculations. SCF calculations are performed with a 
cutoff of $10^{-7}$ for the norm of the error vector. The ground state coupled-cluster calculations uses 
convergence cutoff of $10^{-9}$ and a DIIS space of 6.
The DI-EOMCC method uses convergence threshold of $10^{-5}$.
Scalar real Gaussian functions constitute the finite atomic orbital basis and are used in our calculations in an uncontracted fashion.
Dyall.cv3z\cite{dyall2016relativistic} basis is chosen for Ar.
We opted Dyall.acv3z basis\cite{dyall2006relativistic} for Kr and Xe and dyall.acv4z\cite{dyall2006relativistic} basis for the Rn atom.
For the molecular systems, dyall.acv3z\cite{dyall2016relativistic} basis is chosen for Cl in Cl$_2$, and
dyall.cv3z\cite{dyall2006relativistic} basis
is for Br atom in Br$_2$. 
aug-cc-pVTZ\cite{dunning1989gaussian} basis is chosen for the H atom in both HBr and HI. Dyall.av3z\cite{dyall2006relativistic} basis is used for Br in HBr
and dyall.ae3z\cite{dyall2006relativistic} basis has opted for I in HI.  
We have used experimental bond-length of 1.9870\AA,~ 2.2810\AA,~ 1.4140\AA~, and 1.6090\AA~ for Cl$_2$, Br$_2$, HBr, and HI respectively and 
these values are taken from  the Ref.\cite{huber1979molecular}
\section{results and Discussion}\label{sec4}

\begin{table}[t!]
\caption{Experimental and theoretical DIP values (in eV) in the EOMCC approximation.}
\begin{ruledtabular}
\begin{center}
\begin{tabular}{llrrrr}
System &\,State&\,MBPT(2)&\,2h&\,CCSD&\,Expt.\\
\hline
      
   \\
Cl$_2$ &\,$^{3}\Sigma_{}^{-}$&\, 31.310 &\,35.902 &\,31.397 &\,31.13\cite{cl2_result}\\
       &\,$^{1}\Delta_{}$&\,31.827 &\,36.293 &\,31.907&\,31.74\cite{cl2_result}\\
       &\,$^{1}\Sigma_{}^{+}$&\,33.210&\,36.475&\,32.294&\,32.12\cite{cl2_result}\\
       &\,$^{1}\Sigma_{}^{-}$&\,33.217&\,37.418&\,33.319&\,32.97\cite{cl2_result}\\
\hline       
       \\
Br$_2$ &\,$^{3}\Sigma_{}^{-}$&\,28.752&\,32.380&\,28.473&\,28.53\cite{br2_result}\\
       &\,$^{1}\Delta_{}$     &\,29.327&\,32.935&\,29.041&\,28.91\cite{br2_result}\\
       &\,$^{1}\Sigma_{}^{+}$ &\,29.801&\,33.346&\,29.519&\,29.38\cite{br2_result}\\
       &\,$^{1}\Sigma_{}^{-}$ &\,30.058&\,33.421&\,29.794&\,30.30\cite{br2_result}\\
\hline       
       \\
HBr    &\,$^{3}\Sigma^{-}$       &\,32.688&\,36.859&\,32.757\,&\,32.62 \cite{hbr_result}\\
       &\,$^{1}\Delta$           &\,34.109&\,38.245&\,34.143&\,33.95 \cite{hbr_result}\\
       &\,$^{1}\Sigma^{+}$       &\,35.400&\,38.942&\,35.429&\,35.19 \cite{hbr_result}\\
\hline
HI    &\,$^{3}\Sigma_0^{-}$       &\,29.596&\,33.010&\,29.174\,&\,29.15 \cite{HI_result}\\
       &\,$^{3}\Sigma_1^{-}$       &\,29.837&\,33.399&\,29.412\,&\,29.37 \cite{HI_result}\\
       &\,$^{1}\Delta$           &\,30.931&\,34.458&\,30.481&\,30.39 \cite{HI_result}\\
       &\,$^{1}\Sigma^{+}$       &\,32.238&\,35.367&\,31.801&\,31.64 \cite{HI_result}\\
\end{tabular}
\end{center}
\end{ruledtabular}
\label{mol}
\end{table}
The correlation energies from MBPT(2) and CCSD calculations are reported in Tab. \ref{basis}.
The obtained correlation energies are compared with the values from the DIRAC14\cite{sauedirac}
to test the correctness of the implementation of the ground state calculations. Our results from the correlation calculations match with the DIRAC14 values up to 8-digit after the decimal point.
We have carried out several calculations and found that the agreement is irrespective of the chemical systems or basis sets.
The discrepency beyond this limit is due to the
use of different convergence algorithm or the cutoff used in the storage of the two-body matrix elements.\par
In Tab. \ref{kr_example}, we reported results with various basis sets, and Hamiltonian and taken Kr as an example for comparison.
A two-component description is needed to represent SOC effects.
For the relativistic effective-core-potential (RECP), we have used contracted Gaussian functions,
with point nucleus in a similar fashion as done with scalar-relativistic Hamiltonian with SOC effects.
The calculations using 4-component wavefunction with DC Hamiltonian are also calculated with the same number of virtual orbital in identical conditions.
We have used ECPDS10MDF\cite{peterson2003systematically} relativistic effective-core-potentials (RECP) 
which is optimized at the multiconfiguration Hartree-Fock level
using Dirac-Coulomb Hamiltonian.
It also includes contributions from the Breit interaction.
The spin-free calculation is a one-component in nature.
We have used aug-cc-pVTZ \cite{wilson1999gaussian}  basis for the comparison.
The results justify a clear improvement for 4-component calculations using DC Hamiltonian over the spin-free version of the RECP
and even with the inclusion of spin-orbit interaction.
We have also done a few more calculations with the Dyall basis sets those are uncontracted in nature and specially
designed for the purpose of relativistic calculations. 
The calculations with Dirac-Coulomb or Dirac-Coulomb-Gaunt Hamiltonian with 4-component 
wavefunction uses finite nuclear (Gaussian charge
density distribution) model.
Relativistic solutions have a singularity, and in an approximate treatment (Gaussian basis), it is harder to treat a singularity.
However, a finite nucleus model helps to get rid of the singularity of the orbitals. 
The variations of the DIP values for a given state over the range of chosen basis sets is a maximum of $\sim$0.2 eV.
The contributions of the high-lying orbitals in the computed values beyond 500 Hartree is negligible.
It is also found that the maximum contribution of Gaunt term is about 0.05 eV, though
the current implementation of DIRAC \cite{sauedirac} is includes the Gaunt interaction only in the construction of the Fock matrix and 
transformation of the Gaunt part of the two-electron operator to the molecular orbital (MO) basis is yet to done.
\par 
The numerical results of the lowest five double-ionized states calculated using the four-component relativistic DI-EOMCC method 
by simultaneously removing two-electrons from the closed-shell configuration   
are presented in Tab. \ref{atomic}. Further, the results from the intermediate calculations using MBPT(2) and the $2h$ schemes
are also compiled in the same table.
All these methods are employed to noble gas atoms (Ar, Kr, Xe, and Rn).
We have compared our results with the values from DI-EOMCC calculations with SOC effects. \cite{wang2015equation}
Finally, all these values are tested against the NIST \cite{nist} database.
It is observed that both MBPT(2) and the $2h$ scheme tend to overestimate DI-EOMCC results; however, the deviation is larger for the $2h$ scheme.
$2h$ scheme lacks the contributions of the $3h-1p$ block which is a major source of non-dynamical electron correlation. The ground state 
defined at the MBPT(2) level is rather a better approximation than the $2h$ scheme.
The DI-EOMCC results are found to be very accurate in comparison to the NIST\cite{nist} database for Ar, Kr, and Xe atoms,
and there is a clear improvement over the DI-EOMCC results with SOC effects.\cite{wang2015equation}
The employed virtual subspace for the chosen basis set for the Rn atom is rather small and gives results
similar to the values reported by Wang {\it et al.}  
\cite{wang2015equation}
In their work, the SOC effect is treated perturbatively for the post SCF part using a scalar-relativistic Hamiltonian.
The RECPs are used for taking care of the relativistic effects, and the
one-electron SOC operator is taken from the RECP operator. However, they disregarded the two-electron part of the SOC operator.\cite{dyall2007introduction}
%
Such treatment of the SOC operator
leads to a gross overestimation of spin-orbit
effects. \cite{klein2008perturbative, christiansen2000spin}
There is a large deviation (2.4 eV) from the experimental value reported in the NIST \cite{nist} database.
We take note of the experimental value, which is reported with an uncertainty of 1.9 eV. 
\par
Further, we employed all the three schemes in the DI-EOMCC framework to molecular systems (Cl$_2$, Br$_2$, HBr, and HI) and compared
with the available experimental values.
These results are tabulated in Tab. \ref{mol}.
The reported values for all the states computed using DI-EOMCC scheme with CCSD as a reference
wavefunction are found to be very accurate and relative deviation is well within 2.0\% from the
experimental values. The deviation of the MBPT(2) results are less in comparison to the $2h$ approximation from the experimental
values. Therefore, MBPT(2) is a better approximation than the $2h$ method in the EOMCC framework. 

\section{conclusion}\label{sec5}
We have successfully implemented the relativistic DI-EOMCC method using four-component Dirac spinors
for the calculation of double-ionization spectra. 
The implementation is a general one based on anti-symmetrized
actual two-body matrix elements. It supports both atomic and molecular systems starting
from their closed-shell configuration.
The implemented method is employed to compute a few low-lying doubly-ionized states of both atoms and molecules.
The results of our relativistic DI-EOMCC method found to be very accurate in comparison
to the available experimental values. The outcome of our computation suggests that the MBPT(2)-EOMCC method is a better
approximation than the $2h$-EOMCC scheme. 
\section*{acknowledgments}
Authors acknowledge the resources of the Center of
Excellence in Scientific Computing at CSIR-NCL. K.T.
gratefully acknowledges support from the CSIR for Senior Research fellowship.

\begin{thebibliography}{100}

\bibitem{harris1992double}
F.~Harris,
\newblock International journal of mass spectrometry and ion processes {\bf
  120}, 1 (1992).

\bibitem{vekey1995multiply}
K.~V{\'e}key,
\newblock Mass Spectrometry Reviews {\bf 14}, 195 (1995).

\bibitem{lahmam1989electron}
A.~Lahmam-Bennani, C.~Dupr{\'e}, and A.~Duguet,
\newblock Physical review letters {\bf 63}, 1582 (1989).

\bibitem{ackermann2007operation}
W.~a. Ackermann {\em et~al.},
\newblock Nature photonics {\bf 1}, 336 (2007).

\bibitem{sansone2006isolated}
G.~Sansone {\em et~al.},
\newblock Science {\bf 314}, 443 (2006).

\bibitem{goulielmakis2008single}
E.~Goulielmakis {\em et~al.},
\newblock Science {\bf 320}, 1614 (2008).

\bibitem{cox2003high}
S.~G. Cox {\em et~al.},
\newblock Physical Chemistry Chemical Physics {\bf 5}, 663 (2003).

\bibitem{PhysRevA.51.R2668}
C.~Le~Sech,
\newblock Phys. Rev. A {\bf 51}, R2668 (1995).

\bibitem{prasad1975importance}
S.~Prasad and D.~Furman,
\newblock Journal of Geophysical Research {\bf 80}, 1360 (1975).

\bibitem{rosner1998study}
S.~Rosner, R.~Cameron, T.~Scholl, and R.~Holt,
\newblock Journal of molecular spectroscopy {\bf 189}, 83 (1998).

\bibitem{grant2007relativistic}

\newblock I.~P. Grant{\em Relativistic quantum theory of atoms and molecules:
  theory and computation} Vol.~40 (Springer Science \& Business Media, 2007).

\bibitem{dyall2007introduction}
K.~G. Dyall and K.~F{\ae}gri~Jr,
\newblock {\em Introduction to relativistic quantum chemistry} (Oxford
  University Press, 2007).

\bibitem{bartlett2007coupled}
R.~J. Bartlett and M.~Musia{\l},
\newblock Reviews of Modern Physics {\bf 79}, 291 (2007).

\bibitem{crawford2000introduction}
T.~D. Crawford and H.~F. Schaefer,
\newblock Reviews in computational chemistry {\bf 14}, 33 (2000).

\bibitem{kummel2003biography}
H.~G. K{\"u}mmel,
\newblock International Journal of Modern Physics B {\bf 17}, 5311 (2003).

\bibitem{ilyabaev1992relativistic}
E.~Ilyabaev and U.~Kaldor,
\newblock Chemical physics letters {\bf 194}, 95 (1992).

\bibitem{ilyabaev1992relativistic2}
E.~Ilyabaev and U.~Kaldor,
\newblock The Journal of chemical physics {\bf 97}, 8455 (1992).

\bibitem{ilyabaev1993relativistic}
E.~Ilyabaev and U.~Kaldor,
\newblock Physical Review A {\bf 47}, 137 (1993).

\bibitem{visscher2001formulation}
L.~Visscher, E.~Eliav, and U.~Kaldor,
\newblock The Journal of Chemical Physics {\bf 115}, 9720 (2001).

\bibitem{eliav1995transition}
E.~Eliav, U.~Kaldor, and Y.~Ishikawa,
\newblock Physical Review A {\bf 52}, 2765 (1995).

\bibitem{eliav1996element}
E.~Eliav, U.~Kaldor, Y.~Ishikawa, and P.~Pyykk{\"o},
\newblock Physical review letters {\bf 77}, 5350 (1996).

\bibitem{PhysRevLett.118.023002}
L.~F. Pa\ifmmode~\check{s}\else \v{s}\fi{}teka, E.~Eliav, A.~Borschevsky,
  U.~Kaldor, and P.~Schwerdtfeger,
\newblock Phys. Rev. Lett. {\bf 118}, 023002 (2017).

\bibitem{lindgren1978coupled}
I.~Lindgren,
\newblock International Journal of Quantum Chemistry {\bf 14}, 33 (1978).

\bibitem{haque1984application}
M.~A. Haque and D.~Mukherjee,
\newblock The Journal of chemical physics {\bf 80}, 5058 (1984).

\bibitem{stolarczyk1985coupled}
L.~Z. Stolarczyk and H.~J. Monkhorst,
\newblock Physical Review A {\bf 32}, 725 (1985).

\bibitem{pal1987multireference}
S.~Pal, M.~Rittby, R.~J. Bartlett, D.~Sinha, and D.~Mukherjee,
\newblock Chemical physics letters {\bf 137}, 273 (1987).

\bibitem{lindgren1987connectivity}
I.~Lindgren and D.~Mukherjee,
\newblock Physics Reports {\bf 151}, 93 (1987).

\bibitem{pal1988molecular}
S.~Pal, M.~Rittby, R.~J. Bartlett, D.~Sinha, and D.~Mukherjee,
\newblock The Journal of chemical physics {\bf 88}, 4357 (1988).

\bibitem{jeziorski1989valence}
B.~Jeziorski and J.~Paldus,
\newblock The Journal of Chemical Physics {\bf 90}, 2714 (1989).

\bibitem{mukherjee1989use}
D.~Mukherjee and S.~Pal,
\newblock Use of cluster expansion methods in the open-shell correlation
  problem,
\newblock in {\em Advances in Quantum Chemistry} Vol.~20, pp. 291--373,
  Elsevier, 1989.

\bibitem{malrieu1985intermediate}
J.~Malrieu, P.~Durand, and J.~Daudey,
\newblock Journal of Physics A: Mathematical and General {\bf 18}, 809 (1985).

\bibitem{meissner1998fock}
L.~Meissner,
\newblock The Journal of chemical physics {\bf 108}, 9227 (1998).

\bibitem{landau1999intermediate}
A.~Landau, E.~Eliav, and U.~Kaldor,
\newblock Chemical physics letters {\bf 313}, 399 (1999).

\bibitem{SINHA1989544}
D.~Sinha, S.~Mukhopadhyay, R.~Chaudhuri, and D.~Mukherjee,
\newblock Chemical Physics Letters {\bf 154}, 544  (1989).

\bibitem{sceip}
S.~Chattopadhyay, A.~Mitra, and D.~Sinha,
\newblock The Journal of Chemical Physics {\bf 125}, 244111 (2006).

\bibitem{dyall2016relativistic}
K.~G. Dyall,
\newblock Theoretical Chemistry Accounts {\bf 135}, 128 (2016).

\bibitem{dyall2006relativistic}
K.~G. Dyall,
\newblock Theoretical Chemistry Accounts {\bf 115}, 441 (2006).

\bibitem{dunning1989gaussian}
T.~H. Dunning~Jr,
\newblock The Journal of chemical physics {\bf 90}, 1007 (1989).

\bibitem{wilson1999gaussian}
A.~K. Wilson, D.~E. Woon, K.~A. Peterson, and T.~H. Dunning~Jr,
\newblock The Journal of chemical physics {\bf 110}, 7667 (1999).

\bibitem{peterson2003systematically}
K.~A. Peterson, D.~Figgen, E.~Goll, H.~Stoll, and M.~Dolg,
\newblock The Journal of chemical physics {\bf 119}, 11113 (2003).

\bibitem{nist}
\url{https://physics.nist.gov/PhysRefData/ASD/ionEnergy.html}.

\bibitem{eom1}
H.~Sekino and R.~J. Bartlett,
\newblock Int. J. Quantum Chem. {\bf 26}, 255 (1984).

\bibitem{eom2}
J.~F. Stanton and R.~J. Bartlett,
\newblock J. Chem. Phys. {\bf 98}, 7029 (1993).

\bibitem{eom3}
J.~D. Watts and R.~J. Bartlett,
\newblock Spectrochim. Acta, Part A {\bf 55}, 495 (1999).

\bibitem{eom4}
S.~A. Kucharski, M.~W{\l}och, M.~Musia{\l}, and R.~J. Bartlett,
\newblock J. Chem. Phys. {\bf 115}, 8263 (2001).

\bibitem{eom5}
M.~K{\'a}llay and J.~Gauss,
\newblock The Journal of chemical physics {\bf 121}, 9257 (2004).

\bibitem{eom6}
M.~Musia{\l},
\newblock Mol. Phys. {\bf 103}, 2055 (2005).

\bibitem{eom7}
J.~F. Stanton and J.~Gauss,
\newblock J. Chem. Phys. {\bf 101}, 8938 (1994).

\bibitem{eom8}
J.~F. Stanton and J.~Gauss,
\newblock J. Chem. Phys. {\bf 111}, 8785 (1999).

\bibitem{eom9}
M.~Musia{\l}, S.~A. Kucharski, and R.~J. Bartlett,
\newblock J. Chem. Phys. {\bf 118}, 1128 (2003).

\bibitem{eom10}
M.~Musia{\l} and R.~J. Bartlett,
\newblock Chem. Phys. Lett. {\bf 384}, 210 (2004).

\bibitem{eom11}
M.~Kamiya and S.~Hirata,
\newblock J. Chem. Phys. {\bf 125}, 074111 (2006).

\bibitem{eom12}
M.~Nooijen and R.~J. Bartlett,
\newblock J. Chem. Phys. {\bf 102}, 3629 (1995).

\bibitem{eom13}
M.~Nooijen and R.~J. Bartlett,
\newblock J. Chem. Phys. {\bf 102}, 6735 (1995).

\bibitem{eom14}
M.~Musia{\l} and R.~J. Bartlett,
\newblock J. Chem. Phys. {\bf 119}, 1901 (2003).

\bibitem{eom15}
A.~I. Krylov,
\newblock Annu. Rev. Phys. Chem. {\bf 59}, 433 (2008).

\bibitem{eom16}
J.~R. Gour and P.~Piecuch,
\newblock The Journal of chemical physics {\bf 125}, 234107 (2006).

\bibitem{eom17}
J.~R. Gour, P.~Piecuch, and M.~W{\l}och,
\newblock The Journal of chemical physics {\bf 123}, 134113 (2005).

\bibitem{eom18}
J.~F. Stanton and J.~Gauss,
\newblock The Journal of chemical physics {\bf 103}, 1064 (1995).

\bibitem{eom19}
S.~V. Levchenko and A.~I. Krylov,
\newblock The Journal of Chemical Physics {\bf 120}, 175 (2004).

\bibitem{ueda}
J.~H.~D. Eland {\em et~al.},
\newblock Phys. Rev. Lett. {\bf 105}, 213005 (2010).

\bibitem{nimrod}
L.~S. Cederbaum, Y.-C. Chiang, P.~V. Demekhin, and N.~Moiseyev,
\newblock Phys. Rev. Lett. {\bf 106}, 123001 (2011).

\bibitem{linderberg2004propagators}
J.~Linderberg and Y.~{\"O}hrn,
\newblock Propagators in quantum chemistry, 2004.

\bibitem{cederbaum}
L.~S. Cederbaum, W.~Domcke, and J.~Schirmer,
\newblock Phys. Rev. A {\bf 22}, 206 (1980).

\bibitem{monkhorst1977calculation}
H.~J. Monkhorst,
\newblock International Journal of Quantum Chemistry {\bf 12}, 421 (1977).

\bibitem{mukherjee1979response}
D.~Mukherjee and P.~Mukherjee,
\newblock Chemical Physics {\bf 39}, 325 (1979).

\bibitem{koch1990coupled}
H.~Koch and P.~J{\o}rgensen,
\newblock The Journal of Chemical Physics {\bf 93}, 3333 (1990).

\bibitem{koch1990coupled1}
H.~Koch {\em et~al.},
\newblock The Journal of Chemical Physics {\bf 92}, 4924 (1990).

\bibitem{nakatsuji1978cluster}
H.~Nakatsuji and K.~Hirao,
\newblock The Journal of Chemical Physics {\bf 68}, 2053 (1978).

\bibitem{nakatsuji1979cluster}
H.~Nakatsuji,
\newblock Chemical Physics Letters {\bf 67}, 329 (1979).

\bibitem{nakatsuji1979cluster2}
H.~Nakatsuji,
\newblock Chemical Physics Letters {\bf 67}, 334 (1979).

\bibitem{chaudhuri1999relativistic}
R.~K. Chaudhuri, P.~K. Panda, B.~Das, U.~S. Mahapatra, and D.~Mukherjee,
\newblock Physical Review A {\bf 60}, 246 (1999).

\bibitem{chaudhuri2000ionization}
R.~K. Chaudhuri {\em et~al.},
\newblock Journal of Physics B: Atomic, Molecular and Optical Physics {\bf 33},
  5129 (2000).

\bibitem{dolg2000effective}
M.~Dolg {\em et~al.},
\newblock Modern methods and algorithms of quantum chemistry {\bf 1}, 479
  (2000).

\bibitem{nicklass1995ab}
A.~Nicklass, M.~Dolg, H.~Stoll, and H.~Preuss,
\newblock The Journal of chemical physics {\bf 102}, 8942 (1995).

\bibitem{dolg2012relativistic}
M.~Dolg and X.~Cao,
\newblock Chemical reviews {\bf 112}, 403 (2012).

\bibitem{schwerdtfeger2011pseudopotential}
P.~Schwerdtfeger,
\newblock ChemPhysChem {\bf 12}, 3143 (2011).

\bibitem{hirata2007high}
S.~Hirata, T.~Yanai, R.~J. Harrison, M.~Kamiya, and P.-D. Fan,
\newblock The Journal of chemical physics {\bf 126}, 024104 (2007).

\bibitem{epifanovsky2015spin}
E.~Epifanovsky, K.~Klein, S.~Stopkowicz, J.~Gauss, and A.~I. Krylov,
\newblock The Journal of chemical physics {\bf 143}, 064102 (2015).

\bibitem{asthana2019exact}
A.~Asthana, J.~Liu, and L.~Cheng,
\newblock The Journal of chemical physics {\bf 150}, 074102 (2019).

\bibitem{cao2017coupled}
Z.~Cao, F.~Wang, and M.~Yang,
\newblock The Journal of chemical physics {\bf 146}, 134108 (2017).

\bibitem{cheng2018perturbative}
L.~Cheng, F.~Wang, J.~F. Stanton, and J.~Gauss,
\newblock The Journal of chemical physics {\bf 148}, 044108 (2018).

\bibitem{cao2016spin}
Z.~Cao, F.~Wang, and M.~Yang,
\newblock The Journal of chemical physics {\bf 145}, 154110 (2016).

\bibitem{wang2015equation}
Z.~Wang, S.~Hu, F.~Wang, and J.~Guo,
\newblock The Journal of chemical physics {\bf 142}, 144109 (2015).

\bibitem{wang2014equation}
Z.~Wang, Z.~Tu, and F.~Wang,
\newblock Journal of chemical theory and computation {\bf 10}, 5567 (2014).

\bibitem{tu2012equation}
Z.~Tu, F.~Wang, and X.~Li,
\newblock The Journal of chemical physics {\bf 136}, 174102 (2012).

\bibitem{jensen2017introduction}
F.~Jensen,
\newblock {\em Introduction to computational chemistry} (John wiley \& sons,
  2017).

\bibitem{sakurai1967advanced}
J.~J. Sakurai,
\newblock Advanced quantum mechanics (redwood, 1967.

\bibitem{visser1992relativistic}
O.~Visser, L.~Visscher, P.~Aerts, and W.~Nieuwpoort,
\newblock Theoretica chimica acta {\bf 81}, 405 (1992).

\bibitem{saue2011relativistic}
T.~Saue,
\newblock ChemPhysChem {\bf 12}, 3077 (2011).

\bibitem{pathak2014relativistic}
H.~Pathak, B.~Sahoo, B.~Das, N.~Vaval, and S.~Pal,
\newblock Physical Review A {\bf 89}, 042510 (2014).

\bibitem{blundell2014calculation}
S.~Blundell,
\newblock Physical Review A {\bf 90}, 042514 (2014).

\bibitem{sahoo}
D.~K. Nandy, Y.~Singh, and B.~K. Sahoo,
\newblock Phys. Rev. A {\bf 89}, 062509 (2014).

\bibitem{pathak2015relativistic}
H.~Pathak {\em et~al.},
\newblock Journal of Physics B: Atomic, Molecular and Optical Physics {\bf 48},
  115009 (2015).

\bibitem{pathak2014relativisticmol}
H.~Pathak, S.~Sasmal, M.~K. Nayak, N.~Vaval, and S.~Pal,
\newblock Physical Review A {\bf 90}, 062501 (2014).

\bibitem{pathak2016relativisticea}
H.~Pathak, S.~Sasmal, M.~K. Nayak, N.~Vaval, and S.~Pal,
\newblock Computational and Theoretical Chemistry {\bf 1076}, 94 (2016).

\bibitem{liu2010ideas}
W.~Liu,
\newblock Molecular Physics {\bf 108}, 1679 (2010).

\bibitem{pathak2016relativisticopen}
H.~Pathak, S.~Sasmal, M.~K. Nayak, N.~Vaval, and S.~Pal,
\newblock The Journal of chemical physics {\bf 145}, 074110 (2016).

\bibitem{shee2018equation}
A.~Shee, T.~Saue, L.~Visscher, and A.~Severo Pereira~Gomes,
\newblock The Journal of chemical physics {\bf 149}, 174113 (2018).

\bibitem{nooijen1997similarity}
M.~Nooijen and R.~J. Bartlett,
\newblock The Journal of chemical physics {\bf 107}, 6812 (1997).

\bibitem{nooijen2002state}
M.~Nooijen,
\newblock International Journal of Molecular Sciences {\bf 3}, 656 (2002).

\bibitem{sattelmeyer2003use}
K.~W. Sattelmeyer, H.~F. Schaefer~Iii, and J.~F. Stanton,
\newblock Chemical physics letters {\bf 378}, 42 (2003).

\bibitem{demel2008application}
O.~Demel, K.~Shamasundar, L.~Kong, and M.~Nooijen,
\newblock The Journal of Physical Chemistry A {\bf 112}, 11895 (2008).

\bibitem{musial2011multireference}
M.~Musia{\l}, A.~Perera, and R.~J. Bartlett,
\newblock The Journal of chemical physics {\bf 134}, 114108 (2011).

\bibitem{shen2013doubly}
J.~Shen and P.~Piecuch,
\newblock The Journal of chemical physics {\bf 138}, 194102 (2013).

\bibitem{musial2011multireference1}
M.~Musia{\l}, S.~A. Kucharski, and R.~J. Bartlett,
\newblock Journal of chemical theory and computation {\bf 7}, 3088 (2011).

\bibitem{musial2014equation}
M.~Musia{\l}, {\L}.~Lupa, and S.~A. Kucharski,
\newblock The Journal of chemical physics {\bf 140}, 114107 (2014).

\bibitem{pathak2014relativisticdip}
H.~Pathak {\em et~al.},
\newblock Phys. Rev. A {\bf 90}, 010501 (2014).

\bibitem{shavitt2009many}
I.~Shavitt and R.~J. Bartlett,
\newblock {\em Many-body methods in chemistry and physics: MBPT and
  coupled-cluster theory} (Cambridge university press, 2009).

\bibitem{davidson197514}
C.~Davidson,
\newblock Journal of Computational Physics {\bf 17}, 87 (1975).

\bibitem{sauedirac}
{DIRAC}, a relativistic ab initio electronic structure program, Release
  {DIRAC14} (2014), written by T.~Saue, L.~Visscher, H.~J.~{\relax Aa}.~Jensen,
  and R.~Bast. with contributions from V.~Bakken, K.~G.~Dyall, S.~Dubillard,
  U.~Ekstr{\"o}m, E.~Eliav, T.~Enevoldsen, E.~Fa{\ss}hauer, T.~Fleig,
  O.~Fossgaard, A.~S.~P.~Gomes, T.~Helgaker, J.~K.~L{\ae}rdahl, Y.~S.~Lee,
  J.~Henriksson, M.~Ilia{\v{s}}, Ch.~R.~Jacob, S.~Knecht, S.~Komorovsk{\'y},
  O.~Kullie, C.~V.~Larsen, H.~S.~Nataraj, P.~Norman, G.~Olejniczak, J.~Olsen,
  Y.~C.~Park, J.~K.~Pedersen, M.~Pernpointner, R.~di~Remigio, K.~Ruud,
  P.~Sa{\l}ek, B.~Schimmelpfennig, J.~Sikkema, A.~J.~Thorvaldsen, J.~Thyssen,
  J.~van~Stralen, S.~Villaume, O.~Visser, T.~Winther, and S.~Yamamoto (see
  \url{http://www.diracprogram.org}).

\bibitem{visscher1997dirac}
L.~Visscher and K.~G. Dyall,
\newblock Atomic Data and Nuclear Data Tables {\bf 67}, 207 (1997).

\bibitem{huber1979molecular}
K.~Huber,
\newblock Molecular Spectra and molecular Structure Constants of Diatomic
  molecules  (1979).

\bibitem{cl2_result}
A.~McConkey {\em et~al.},
\newblock Journal of Physics B: Atomic, Molecular and Optical Physics {\bf 27},
  271 (1994).

\bibitem{br2_result}
T.~Fleig, D.~Edvardsson, S.~T. Banks, and J.~H. Eland,
\newblock Chemical Physics {\bf 343}, 270 (2008).

\bibitem{hbr_result}
J.~H. Eland,
\newblock Chemical physics {\bf 294}, 171 (2003).

\bibitem{HI_result}
A.~J. Yencha {\em et~al.},
\newblock Chemical physics {\bf 303}, 179 (2004).

\bibitem{klein2008perturbative}
K.~Klein and J.~Gauss,
\newblock The Journal of chemical physics {\bf 129}, 194106 (2008).

\bibitem{christiansen2000spin}
O.~Christiansen, J.~Gauss, and B.~Schimmelpfennig,
\newblock Physical Chemistry Chemical Physics {\bf 2}, 965 (2000).

\end{thebibliography}



\end{document}